\documentclass[runningheads]{llncs}
\usepackage[T1]{fontenc}

\usepackage{graphicx}
\usepackage{amsmath}
\usepackage{graphicx}
\usepackage{epstopdf}

\usepackage{xcolor}
\definecolor{lightgrey}{rgb}{0.9, 0.9, 0.9}
\usepackage{tabularray}
\usepackage[ruled,vlined]{algorithm2e}
\usepackage{caption}
\usepackage{subcaption}
\usepackage{amssymb}
\usepackage{amsbsy}
\usepackage{rotating}

\usepackage[normalem]{ulem}

\usepackage{fancyhdr}
\definecolor{commentcolor}{RGB}{110,154,155}   

\begin{document}
\title{MAR-DTN: Metal Artifact Reduction using Domain Transformation Network for Radiotherapy Planning}

\titlerunning{MAR-DTN}
%
\author{Belén Serrano-Antón\inst{1,2,3}\orcidID{0000-0001-9328-9525} \and
Mubashara Rehman\inst{4,5}\orcidID{0009-0007-2935-0409}\thanks{co-first author of the manuscript} \and
Niki Martinel\inst{4}\orcidID{0000-0002-6962-8643} \and
Michele Avanzo\inst{5}\orcidID{0000-0003-1711-4242} \and
Riccardo Spizzo\inst{5}\orcidID{0000-0001-7772-0960} \and
Giuseppe Fanetti\inst{5}\orcidID{0000-0002-9677-3176} \and
Alberto P. Muñuzuri\inst{2,3}\orcidID{0000-0002-0579-9347}\and
Christian Micheloni\inst{4}\orcidID{0000-0003-4503-7483}}
\authorrunning{SA.Belén, R.Mubashara et al.}


%
\institute{FlowReserve Labs S.L., 15782 Santiago de Compostela, Spain 
\and CITMAga, 15782 Santiago de Compostela, Spain 
\and Group of Nonlinear Physics, University of Santiago de Compostela, 15782 Santiago de Compostela, Spain 
\and Machine Learning and Perception Lab, Università degli Studi di Udine, 33100 Udine (UD) Italy
\and Centro di Riferimento Oncologico di Aviano IRCCS, 33081 Aviano (PN) Italy}



%
\maketitle              
\begin{abstract}


For the planning of radiotherapy treatments for head and neck cancers, Computed Tomography (CT) scans of the patients are typically employed. However, in patients with head and neck cancer, the quality of standard CT scans generated using kilo-Voltage (kVCT) tube potentials is severely degraded by streak artifacts occurring in the presence of metallic implants such as dental fillings. Some radiotherapy devices offer the possibility of acquiring Mega-Voltage CT (MVCT) for daily patient setup verification, due to the higher energy of X-rays used, MVCT scans are almost entirely free from artifacts making them more suitable for radiotherapy treatment planning.

In this study, we leverage the advantages of kVCT scans with those of MVCT scans (artifact-free). We propose a deep learning-based approach capable of generating artifact-free MVCT images from acquired kVCT images.
The outcome offers the benefits of artifact-free MVCT images with enhanced soft tissue contrast, harnessing valuable information obtained through kVCT technology for precise therapy calibration. Our proposed method employs UNet-inspired model, and is compared with adversarial learning and transformer networks. This first and unique approach achieves remarkable success, with PSNR of $30.02$dB across the entire patient volume and $27.47$dB in artifact-affected regions exclusively.  It is worth noting that the PSNR calculation excludes the background, concentrating solely on the region of interest.

\end{abstract}
%
%
\section{Introduction}
Since their introduction in the 1970s, advanced medical imaging techniques, particularly high-resolution Computed Tomography (CT), have been crucial for computer-assisted diagnosis \cite{10.1259/0007-1285-46-552-1016}. However, when patients with metal implants undergo imaging, such as dental fillings or hip prostheses, severe beam attenuation occurs, resulting in discernible streaks that compromise image fidelity and hampering clinical assessment \cite{Boas2012article}.

Recent advancements in deep learning have shown promise in mitigating metal artifacts through supervised learning methodologies. However, obtaining ground truth images without artifacts is challenging.
\cite{KAPOSI202078.e17} tackles this issue by generating datasets with and without metal artifacts, enabling the development of numerous algorithms for Metal Artifact Reduction (MAR). 
Other approaches encompass a variety of image-to-image deep learning models, including deep residual architectures \cite{huang2018metal} and interpretable convolutional dictionary networks \cite{wang2021dicdnet}. 
Numerous other methodologies utilize sinogram-to-sinogram deep learning models \cite{https://doi.org/10.1002/mp.13199,s21248164} or dual-domain deep learning models using both image and sinogram data \cite{8953298,LeePLC20,yu2021metal}. These models can be further extended by incorporating state-of-the-art interpolation-based algorithm Normalized MAR corrected data as an extra input \cite{gjesteby2019dual,liang2019metal}. A combination of multiple supervised deep learning methods can be effective in reducing metal artifacts from complex cases of cardiac CT images \cite{LOSSAUNEEELSS2020101655}. Approach \cite{wang2018conditional} uses  pix2pix \cite{8100115} for MAR, it introduces band-wise normalization method, which splits a CT image into three channels according to the intensity value and considerably improves the performance of the cGAN. CNN-based approach \cite{8331163} is introduced to predict an artifact-suppressed prior image. Extending these concepts, \cite{8953298} introduced DuDoNet, a dual-domain learning technique combining sinogram enhancement and image domain reconstruction. Improved version of DuDoNet \cite{8953298}, restores sinogram consistency and simultaneously enhance CT images by incorporating metal segmentation in both domains. In more recent work, \cite{10.1007/978-3-030-59713-9_15} introduced an alternative dual-domain approach, emphasizing deep sinogram completion for improved MAR performance. 


Mega-Voltage Computed Tomography (MVCT) is used for verification of patient positioning immediately before the radiotherapy treatment. It is less prone to streak artifacts from metallic implants because it uses high-energy beams produced by a radiotherapy linear accelerator, which are less attenuated by metal than conventional diagnostic X-rays. The main drawback of MVCT is that it is available only in some specialized radiotherapy machines \cite{10.1259/0007-1285-46-552-1016}. \cite{liugang2016metal} proposes to reduce metal artifacts in kVCT by using MVCT images as prior images. The iterative method proposed in~\cite{ni2023metal} segments tissue regions in Megavoltage cone-beam CT images and the metal region in kVCT images for template creation. Forward projection of the templates generates sinograms. Artifact images are reconstructed from the sinograms. Finally, corrected images are obtained by subtracting artifact images from original kVCT images.~\cite{paudel2014clinical} utilizes the sinogram of kVCT and MVCT along with the corresponding metal trace to ultimately produce artifact-free kVCT images. Methodology proposed in \cite{kim2022metal}, employing convolutional neural networks to obtain artifact-free kVCT images, by utilizing two networks where the first generates synthetic artifact-free kVCT images from MVCT, which are then used to train the second network. The second network takes kVCT images with artifacts as input and produces artifact-free kVCT images as output.

\begin{figure}[h]
    \centering

    \begin{subfigure}[b]{\textwidth}
        \centering
        \includegraphics[width=1.0\textwidth]{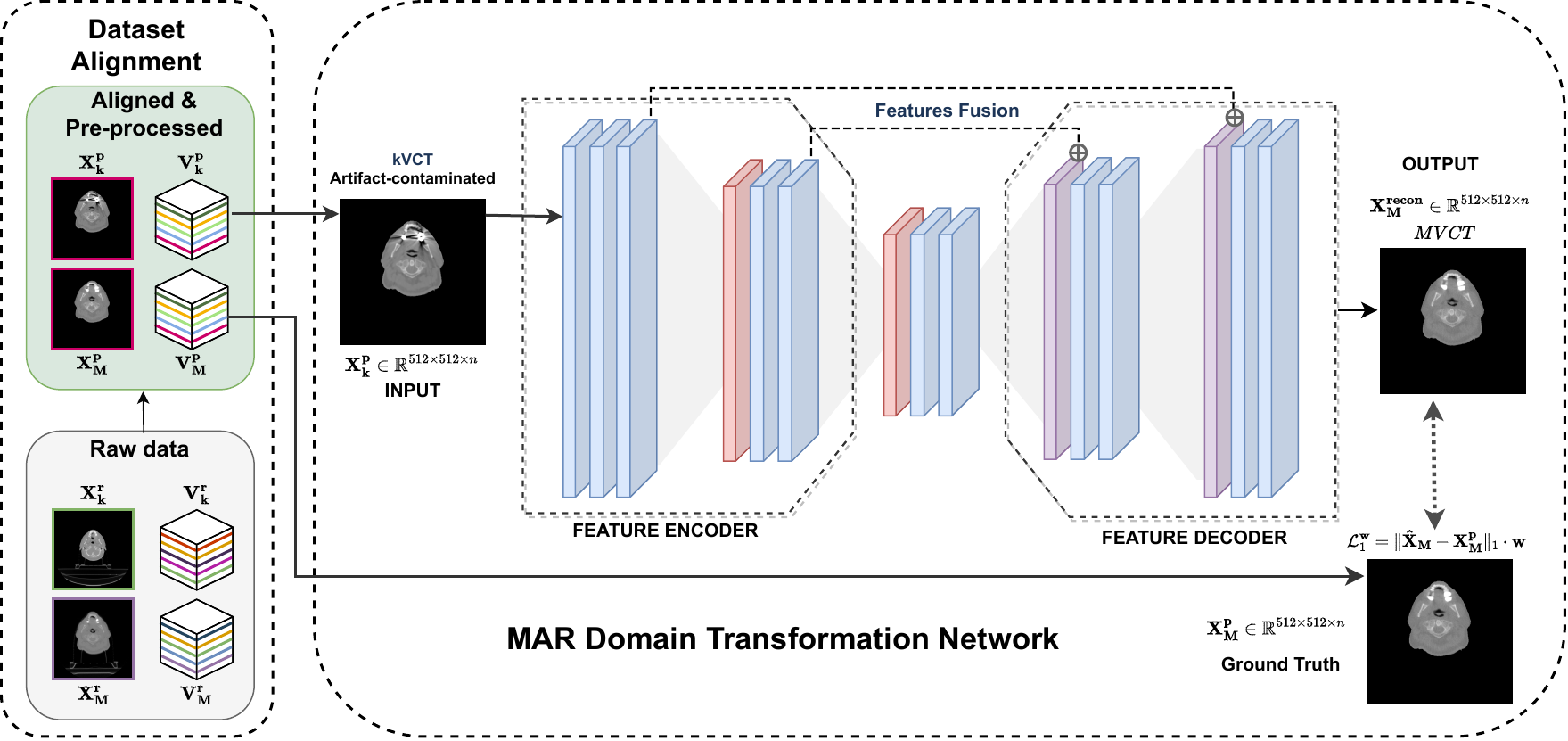}
        \caption{}
        \label{fig:network}
    \end{subfigure}
    
    \begin{subfigure}[b]{0.4\textwidth}
        \centering
        \includegraphics[width=0.5\textwidth]{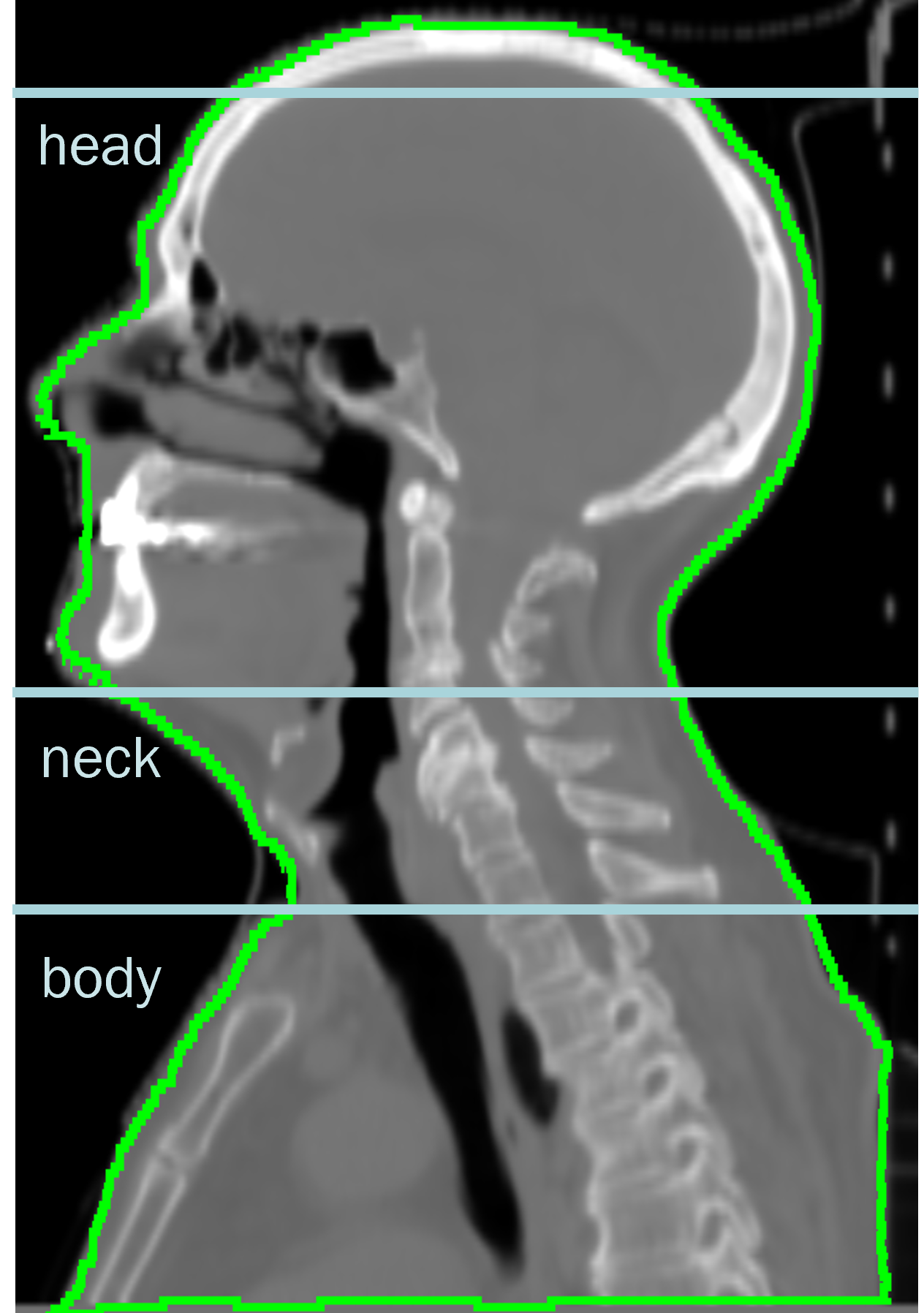}
        \caption{}
        \label{fig:regionsAndBodySegmenation}
    \end{subfigure}
    \hfill
    \begin{subfigure}[b]{0.4\textwidth}
        \centering
        \includegraphics[width=0.5\textwidth]{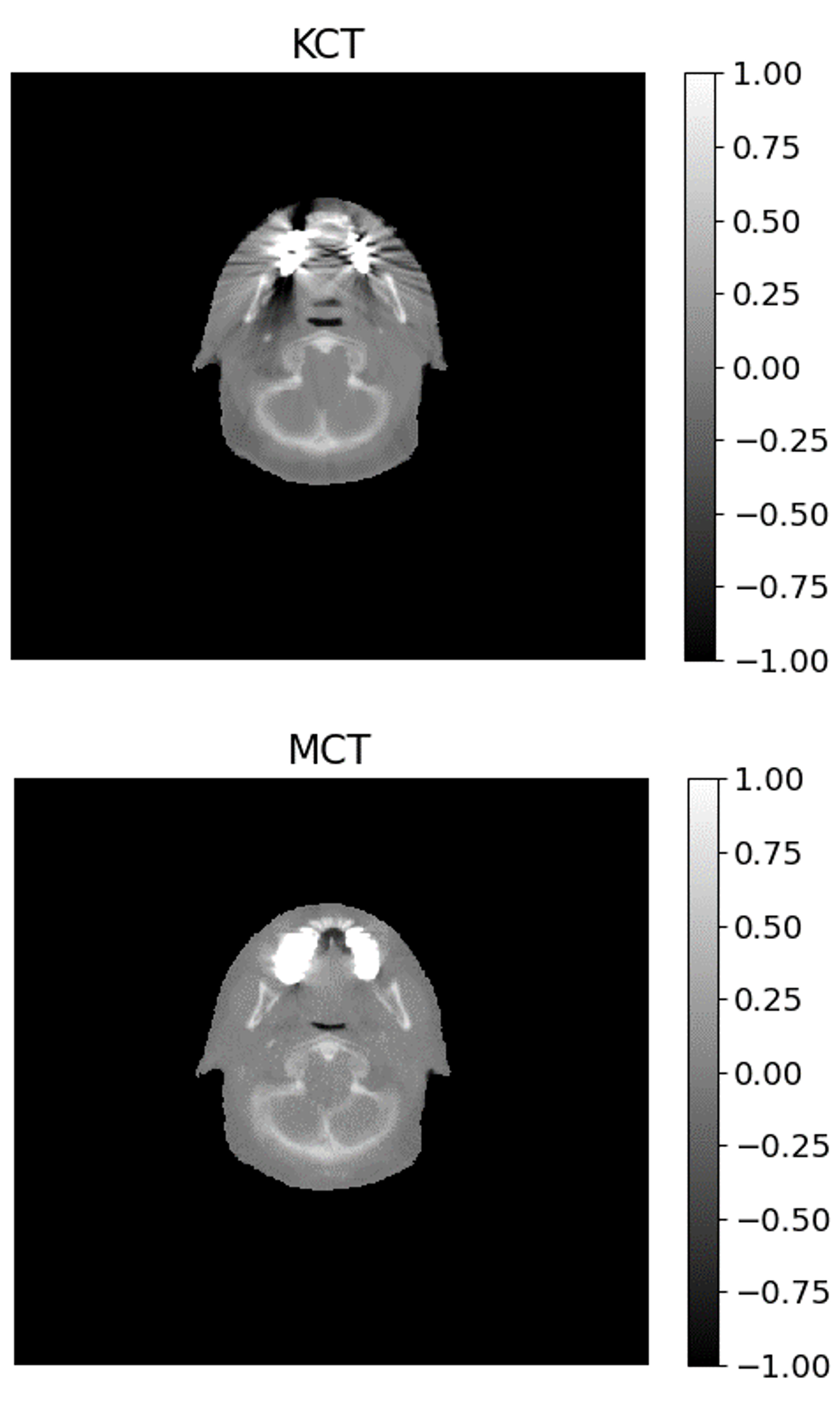}
        \caption{}
        \label{fig:normalizationVertical}
    \end{subfigure}

    \caption{(a) Abstract overview of proposed Domain Transformation Network. (b) Sagittal view of the body with distinct delineations of the head, neck, and body regions(blue). (c) kVCT (top) and MVCT (bottom) axial artifact slices after normalization and masking.}
    \label{fig:regionsAndPrepro}
\end{figure}

Different techniques have been utilized to mitigate metal artifacts in kVCT scans, yet they predominantly operate within the same domain. In contrast, our innovative approach involves transforming the CT domain from kVCT to MVCT, as MVCT is inherently less artifact-sensitive, henceforth preferred for its robustness in clinical applications. We hereby propose Metal Artifact Reduction using Domain Transformation Network (MAR-DTN) to address metal artifacts in oncological imaging.
Our approach generates MVCT using a model that employs a UNet architecture with skip connections, tailored for MAR from kVCT images, to systematically mitigate artifacts during the transformation process from kVCT to MVCT. Leveraging its encoder-decoder structure and spatial awareness, it effectively processes 512x512 pixel images to produce MVCT output (see Fig.  \ref{fig:network}). This network is trained by employing $3858$ kVCT slices of head and neck region (Fig. \ref{fig:regionsAndBodySegmenation}).

It has achieved exceptional results, a noticeable point is that Peak Signal-to-Noise Ratio (PSNR) and Structural Similarity Index Measure (SSIM) in all tables concern the regions of interest only, not the background like the other methodologies. This approach aims to improve CT quality, interpretability, and analysis of medical images by transforming artifact-contaminated kVCT into artifact-free MVCT. This novel method enables radiation oncologists to gain insights into MVCT from kVCT alone, potentially avoiding repeated imaging and its implications for patients' health.  In addition, our study entails a comparative evaluation of MAR-DTN's performance in relation to three current state-of-the-art methods, based on adversarial learning and transformers.

\section{Methods}
\subsection{Dataset Collection and Processing}
\label{sub:dataset}
Due to the lack of available aligned kVCT and MVCT datasets, a new dataset consisting of $5469$ images from $52$ patients from the \textit{National Cancer Institute (CRO) IRCCS}\footnote{Centro di Riferimento Oncologico di Aviano IRCCS, Via F.Gallini 2, Aviano (PN), 33081, Italy}. For each patient, we acquired kVCT and MVCT images; the kVCT images obtained have matrix size of $512\times512$, on the axial plane with a pixel size of $1.074$ mm $\times$ $1.074$ mm, and slice thickness of $2$ mm, furthermore, the MVCT images obtained have a matrix size of $512\times512$, on the axial plane with a pixel spacing of $0.754$ mm $\times$ $0.754$ mm with the slice thickness of $2$ mm or $4$ mm.

Patients underwent intensity-modulated radiotherapy for oropharyngeal or nasopharyngeal cancer. Non-contrast-enhanced CT imaging was performed using a 32-slice scanner (Toshiba Aquilion LB, Toshiba Medical Systems Europe, Zoetermeer, the Netherlands) with parameters set at 120 kVp, 2-5 mm slice thickness, and 1.07-1.17 mm pixel size. Additionally, patients underwent scanning with helical tomotherapy (Hi-Art II Tomotherapy System, Tomotherapy Inc., Madison, Wisconsin, USA), utilizing a radiotherapy 6MV linear accelerator capable of acquiring MVCT images for daily patient setup verification. The imaging beam, produced by the same LINAC as the therapeutic beam, had a nominal energy of 3.5 MV, with slice thickness ranging from 2-5 mm and a pixel size of 0.75 mm.

The slices of each modality volume are manually categorized into three regions: head, neck, and body (see Fig.  \ref{fig:regionsAndBodySegmenation}). The head region comprises from the beginning of the cranial cavity to the chin, while the neck region spans from the chin to the shoulders.
The remaining (body region) slices are not considered since we care about removing artifacts caused by metal implants in the teeth area.
To separate the artifact-corrupted slices from artifact-free slices, we define artifacts in kVCT images as values exceeding $2000$ Hounsfield Units (HU), while for MVCT images, the artifact threshold is set at $1000$ HU.
These thresholds were determined through visual inspection and following recommendations from~\cite{kim2022metal,liugang2016metal}.

\begin{table}[h]
\centering
\caption{Number of patients and slices (images) in the acquired dataset.The head and neck region include the artifact slices since we work with artifacts caused by metallic dental implants.}
\label{table:dataset}
\begin{tblr}{
  cells = {c},
  hline{1,5} = {-}{0.12em},
  hline{2} = {-}{},
}
Set        & Number of patients & {Number of slices of the \\head and neck regions} & {Number of slices \\ with artifacts} \\
Train      & $36$               & $3858$                                            & $560$                                \\
Validation & $10$               & $1031$                                            & $153$                                \\
Test       & $6$                & $580$                                             & $96$                                 
\end{tblr}
\end{table}


For the training and subsequent evaluation of the proposed model, two datasets are constructed; the first is $\mathcal{D}_\text{All}$, the dataset comprises CT slices up to the neck region (including slices with and without artifacts), and the second dataset is $\mathcal{D}_\text{Art}$, which contains only artifact-contaminated CT slices. Out of the total number of slices in the dataset, 14.78\% exhibit artifacts, hence belong to $\mathcal{D}_\text{Art}$. Both datasets are further sub-divided into three distinct datasets, as specifically, $70\%$ of the patients are used for training (\(\mathcal{D}_\text{All}^{Tr}\) and \(\mathcal{D}_\text{Art}^{Tr}\)), $20\%$ for validation (\(\mathcal{D}_\text{All}^{Val}\) and \(\mathcal{D}_\text{Art}^{Val}\)), and the remaining $10\%$ for testing (\(\mathcal{D}_\text{All}^{Ts}\) and \(\mathcal{D}_\text{Art}^{Ts}\)) (see Table \ref{table:dataset}).

\subsection{kVCT-MVCT Alignment and Preprocessing}
\label{sub:alignmentPreprocessing}

The primary goal is to create a dataset with aligned kVCT and MVCT images. Despite originating from the same patient and reference system (with the same origin point), both image volumes (kVCT and MVCT) were not pixel-aligned leading to increased challenges (\textit{i.e.}, such as the need to address alignment and artifact reduction simultaneously).
To achieve this, image alignment was performed using the Elastix module of $3$D Slicer, open source software (version $5.6.1$) \cite{fedorov20123d,klein2009elastix}.

The aligned kVCT and MVCT volumes undergo normalization to the range $[-1,1]$. This process involves setting the lower threshold at $-1000$ for air and upper thresholds at $2000$ for kVCT artifacts and $1000$ for MVCT artifacts. Additionally, utilizing the segmentation provided by clinicians (depicted as green segmentation in Fig.  \ref{fig:regionsAndBodySegmenation}), the image background is standardized to the value of $-1$. The result of such a preprocessing on two sample kVCT and MVCT slices is shown in~\figurename\ref{fig:normalizationVertical}.

\subsection{Proposed Methodology}
 
The objective is to project images acquired in the kVCT domain onto the MVCT domain while removing/reducing the artifacts induced by metallic implants.

In what follows, with $\mathbf{m}$ denoting the kVCT ($\mathbf{k}$) or MVCT ($\mathbf{M}$) modality, we let $\mathbf{X_m^r} \in \mathbb{R}^{d \times d}$ be a raw ($\mathbf{r}$) slice with $d=512$ denoting the image resolution.
The volume containing the $n$ slices of a patient is $\mathbf{V_m^r} \in \mathbb{R}^{d \times d \times n}$.
The original images undergo an alignment process (see Section \ref{sub:alignmentPreprocessing}), resulting in two new volumes, $\mathbf{V_k^a}, \mathbf{V_M^a} = alignment(\mathbf{V_k^r}, \mathbf{V_M^r})$, which are aligned pixel by pixel. 

\begin{figure}[h]
\centering
\includegraphics[width=0.8\textwidth]{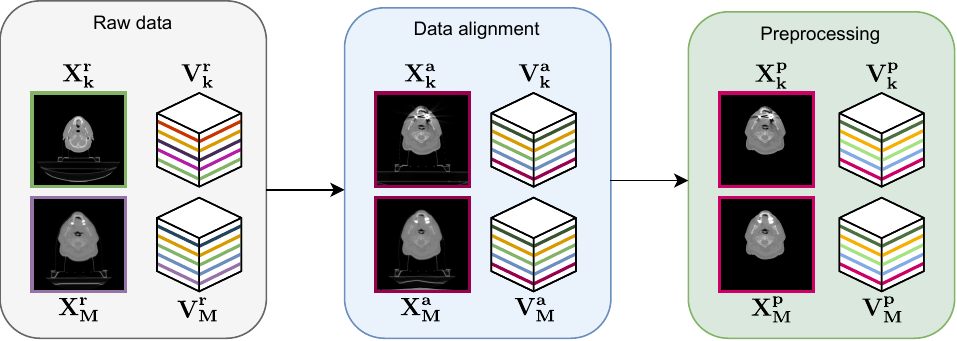}
\caption{Steps followed for dataset generation. We start with raw and unaligned kVCT and MVCT volumes --slices (lines in the cube) do not correspond.
Then, volumes are pixel-aligned and so the slices correspond (Section \ref{sub:alignmentPreprocessing}). 
Finally, corresponding slices in kVCT and MVCT volumes are normalized and masked (Section \ref{sub:alignmentPreprocessing}).} 
\label{fig:rawAlignPrepro}
\end{figure}

After this process, all the slices in a volume are preprocessed (see Section \ref{sub:alignmentPreprocessing}) to obtain $\mathbf{X_m^p} = preprocess(\mathbf{X_m^a}), \forall \mathbf{X_m^a} \in \mathbf{V_m^a}$ that collectively define the dataset for the experiments.
The summary diagram is shown in~\figurename\ref{fig:rawAlignPrepro}.

The input to our model is a preprocessed kVCT image, $\mathbf{X_k^p}$, while the ground truth is the corresponding preprocessed MVCT image, $\mathbf{X_M^p}$.
The output of the model is the domain transferred kVCT to MVCT slice, denoted as $\mathbf{\hat{X}_M} \in \mathbb{R}^{d\times d}$. 

\subsubsection{Network architectures}

We propose a Metal Artifact Reduction using Domain Transformation Network (MAR-DTN), which closely aligns with the architectural principles of the UNet framework \cite{ronneberger2015u}. The UNet architecture has been widely used in previous works for pixel-to-pixel image tasks. In medical imaging, specifically, it has demonstrated excellent results in segmentation, denoising, and MAR \cite{ronneberger2015u,serrano2023coronary,wang2018conditional}. The detailed architectural explanation of our proposed model, named MAR-DTN, can be found in the supplementary material (Section 1: Explanation of Proposed Model Architecture).

Our investigation involves a comparative analysis of the performance of MAR-DTN against three contemporary state-of-the-art methods. The first one is a Conditional Generative Adversarial Network (cGAN), named pix2pix~\cite{8100115}.

Additionally, a modification of this network is included, replacing the original generator with MAR-DTN (referred to as custom-pix$2$pix).

In addition, a network leveraging transformers is implemented due to their demonstrated high performance in addressing pixel-to-pixel image tasks. The SwinIR architecture \cite{liang2021swinir} is structured into three key components: shallow feature extraction, deep feature extraction, and high-quality image reconstruction. Notably, the deep feature extraction module integrates numerous residual Swin Transformer blocks (RSTB), each incorporating multiple Swin Transformer layers alongside a residual connection.

Finally, an architecture initially designed for medical image segmentation is included. This is the INet architecture, a network that does not perform downsampling. It simply enlarges receptive fields by increasing the kernel sizes of convolutional layers in steps (e.g., from 3 × 3 to 7 × 7 and then 15 × 15). In our case, the final activation is not performed in order to obtain a network capable of generating images. We used this architecture for image generation because INet maintains spatial information by fixing the sizes of feature maps and fuses multilevel semantics by concatenating feature maps of all preceding layers. This allows INet to enhance optimization capabilities.

\subsubsection{Loss functions}
\label{subsub:lossFunc}
In addressing artifact reduction with neural networks, various loss functions such as L$1$, FFL (Focal Frequency Loss), MSE (Mean Squared Error), SSIM (Structural Similarity Index), and MS-SSIM (Multi-Scale Structural Similarity Index) offer distinct advantages. 

\begin{itemize}
    \item The weighted $\mathcal{L}_1^\mathbf{w}$ loss function is defined by: $\mathcal{L}_1^\mathbf{w} = \lVert \mathbf{\hat{X}_M} -  \mathbf{X_M^p}\rVert_1 \cdot \mathbf{w}$, where $\mathbf{w} \in \mathbb{R}^{d \times d}$ is the pixel weight. This loss emphasizes the absolute differences between predicted and ground truth values and penalizes outliers, contributing to robust artifact reduction.

    \item FFL \cite{jiang2021focal} is defined by: $\mathcal{L}_{\textit{FFL}}^{\beta,\alpha} = \frac{1}{d \cdot d} \sum_{u=0}^{d-1}\sum_{v=0}^{d-1} z(u,v) |F_{\mathbf{\hat{X}_M}}(u,v) - F_{\mathbf{X_M^p}}(u,v)|^2 \cdot \beta,$ where $z(u,v) = |F_{\mathbf{\hat{X}_M}}(u,v) - F_{\mathbf{X_M^p}}(u,v)|^\alpha$, $\beta \in \mathbb{R}$ is the weight of spatial frequency, $\alpha \in \mathbb{R}$ is the scaling factor,  and $F(u,v)$ is the spatial frequency value at the spectrum coordinate $(u,v)$. This loss focuses on high-frequency artifacts, helps in preserving image details while suppressing artifacts, thus enhancing perceptual quality.

    \item MSE is defined by $\| \mathbf{\hat{X}_{M_{ij}}} -  \mathbf{X_{M_{ij}}} \|^2_2$
    which measures the average squared distance, and provides simplicity and ease of interpretation.

    \item SSIM~\cite{wang2004image}, evaluates luminance, contrast and structure, ensuring preservation of perceptual features, making them suitable for maintaining image fidelity during artifact reduction tasks. 
    


    \item MS-SSIM divides images into multiple scales and computes SSIM for each scale separately. Then, it averages these SSIM values to get a single value representing structural similarity. This method offers a more comprehensive evaluation, considering structural similarity across different resolutions.
        
\end{itemize}

\subsubsection{Implementation details and evaluation metrics}
All networks have the same input and output shape, $512\times512$, corresponding to the size of $\mathbf{X_k^p}$ and $\mathbf{X_M^p}$.
Models were optimized using Adam with learning rate and weight decay set to $0.001$ and $5e^{-4}$, respectively.
The batch size was set to $4$ for all networks except SwinIR for which we used $2$ samples (due to computational memory issues).
We trained for $20$ epochs with early stopping with a patience of $5$ epochs.
Data augmentation~\cite{Buslaev_2020} includes horizontal flip with a probability of $0.5$ and shift, scale, and rotate with a probability of $0.8$ (\textit{shift\_limit}$=0.0625$, \textit{scale\_limit}$=0.1$, \textit{rotate\_limit}$=5$). This introduces variability into our dataset by applying transformation probabilities to alter the dataset in each epoch, thus aiding in the mitigation of data limitation.

Models were trained on an Intel Xeon Server with $188$GB of RAM and an Nvidia A$100$ GPU. We evaluated our methodology using PSNR and SSIM metrics.

\section{Experimental Results}

\subsection{Loss function analysis}
The impact of different loss functions, whether used individually or in combination, is analyzed in this study. We excluded the INet network from our evaluation because its performance, as detailed in Section \ref{sub:lossFuncResults}, is significantly lower compared to the other architectures. Including INet could skew the comparative analysis and potentially introduce biases, thus detracting from a fair assessment of the loss functions' effects on more competitive networks.

First, we explore the impact of using an $L_1$ loss function with weights ($\mathcal{L}_1^\mathbf{w}$) on images containing artifacts. Weight assignment is based on body segmentation provided by clinicians (see Fig.  \ref{fig:regionsAndBodySegmenation}), where $\mathbf{w}[i,j]$ is set to $0.1$ outside the body segment and varies within the set $\{1, 25, 50, 100\}$ inside the segment for slices with artifacts. Slices without artifacts maintain a weight of $1$ throughout the body segment. Since the only variable is the weight within the body segment, we simplify the notation in the following sections and denote this value as $w$. Therefore, $\mathcal{L}_1^{100}$ indicates a weight of $100$ within the body segment for slices with artifacts.

Additionally, the parameters \textit{$\beta$} and \textit{$\alpha$} of the $\mathcal{L}_{\textit{FFL}}^{\beta,\alpha}$ are discretely varied in the set of values $\{0.5,1,1.5\}$. This variation allows for exploring different weightings and contributions of both parameters in the neural network's learning process, particularly in handling images with artifacts.

\subsubsection{$\mathcal{L}_1^\mathbf{w}$ Analysis}

\begin{figure}[htbp]
    \centering

    \begin{subfigure}[b]{0.49\textwidth}
        \centering
        \includegraphics[width=0.77\textwidth]{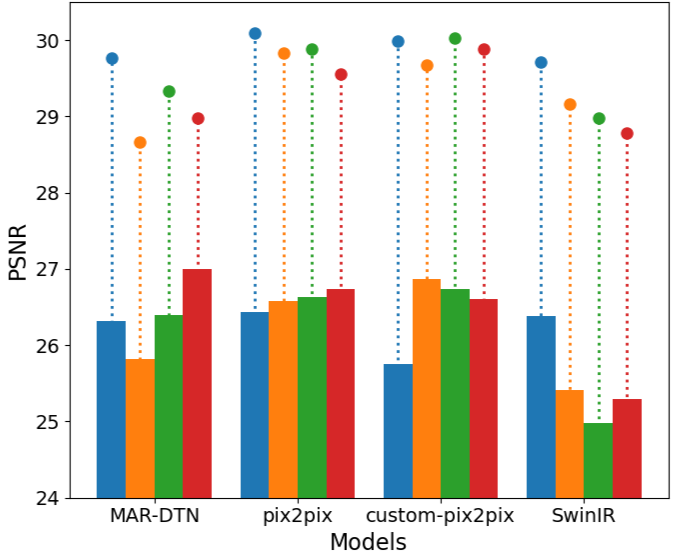}
        \caption{}
        \label{fig:ab_combined_ssim_psnr_weight_a}
    \end{subfigure}
    \hfill
    \begin{subfigure}[b]{0.49\textwidth}
        \centering
        \includegraphics[width=0.8\textwidth]{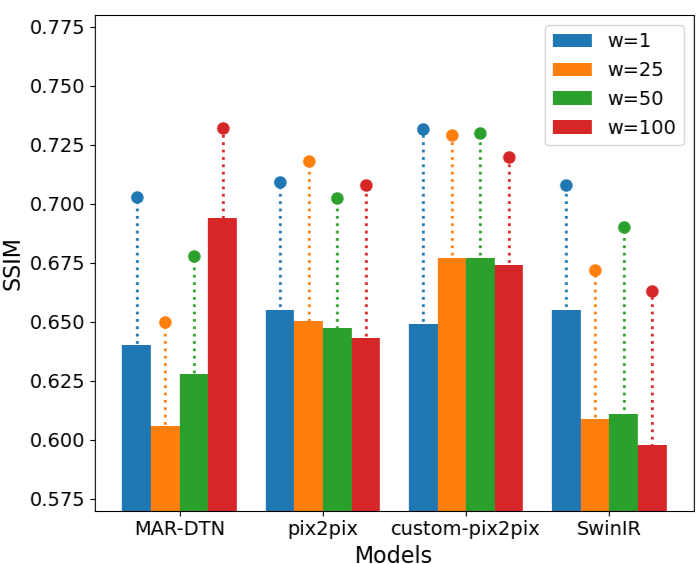}
        \caption{}
        \label{fig:ab_combined_ssim_psnr_weight_b}
    \end{subfigure}

    \caption{PSNR (a) and SSIM (b) values evaluated on the \(\mathcal{D}_\text{All}\). The dots represent the mean value of all slices in the dataset, while the bars represent the mean value of slices with artifacts. Values obtained using the four considered networks (MAR-DTN, pix$2$pix, custom-pix$2$pix and SwinIR) trained on the \(\mathcal{D}_\text{All}\) with the $\mathcal{L}_1^{w}$ loss function only. 
    }
    \label{fig:new}
\end{figure}

Fig. \ref{fig:ab_combined_ssim_psnr_weight_a} and Fig. \ref{fig:ab_combined_ssim_psnr_weight_b} show the PSNR and SSIM values obtained by the networks of the study after training with $\mathcal{L}_1^w$ loss when $w \in \{1,25,50,100\}$ using \(\mathcal{D}_\text{All}\).

The first thing to note is the limited variability of results obtained when modifying the parameter $w$. In terms of PSNR, the results do not vary by more than $3$dB, while for SSIM, the results demonstrate a variance of no more than $10\%$.

In the case of MAR-DTN, a positive trend in the artifact set is observed when $w>25$ increases. Conversely, with SwinIR, better results are achieved when no supplementary weight is allocated to the artifact class. Moreover, when $w>1$ parameters have similar results. 

For pix$2$pix, no significant difference in results is observed, and the same holds true for custom-pix$2$pix when $w>1$. However, increasing the value of \textit{w} for the artifact class does lead to an improvement in the artifact set results.

Regarding the results in the \(\mathcal{D}_\text{All}\), represented by dots in Fig.  \ref{fig:ab_combined_ssim_psnr_weight_a} and Fig. \ref{fig:ab_combined_ssim_psnr_weight_b}, we observe slightly inferior results when giving more weight to the \(\mathcal{D}_\text{Art}\) set. However, we are not concerned as the focus is on the artifact region. Therefore, for the remaining experiments, we will use $w=100$, as it yields the best result for MAR-DTN and similar values to the state-of-the-art $L_1$ for the rest of the networks.

\begin{figure}[htbp]
    \centering
    \begin{subfigure}[b]{0.49\textwidth}
        \centering
        \includegraphics[width=0.9\textwidth]{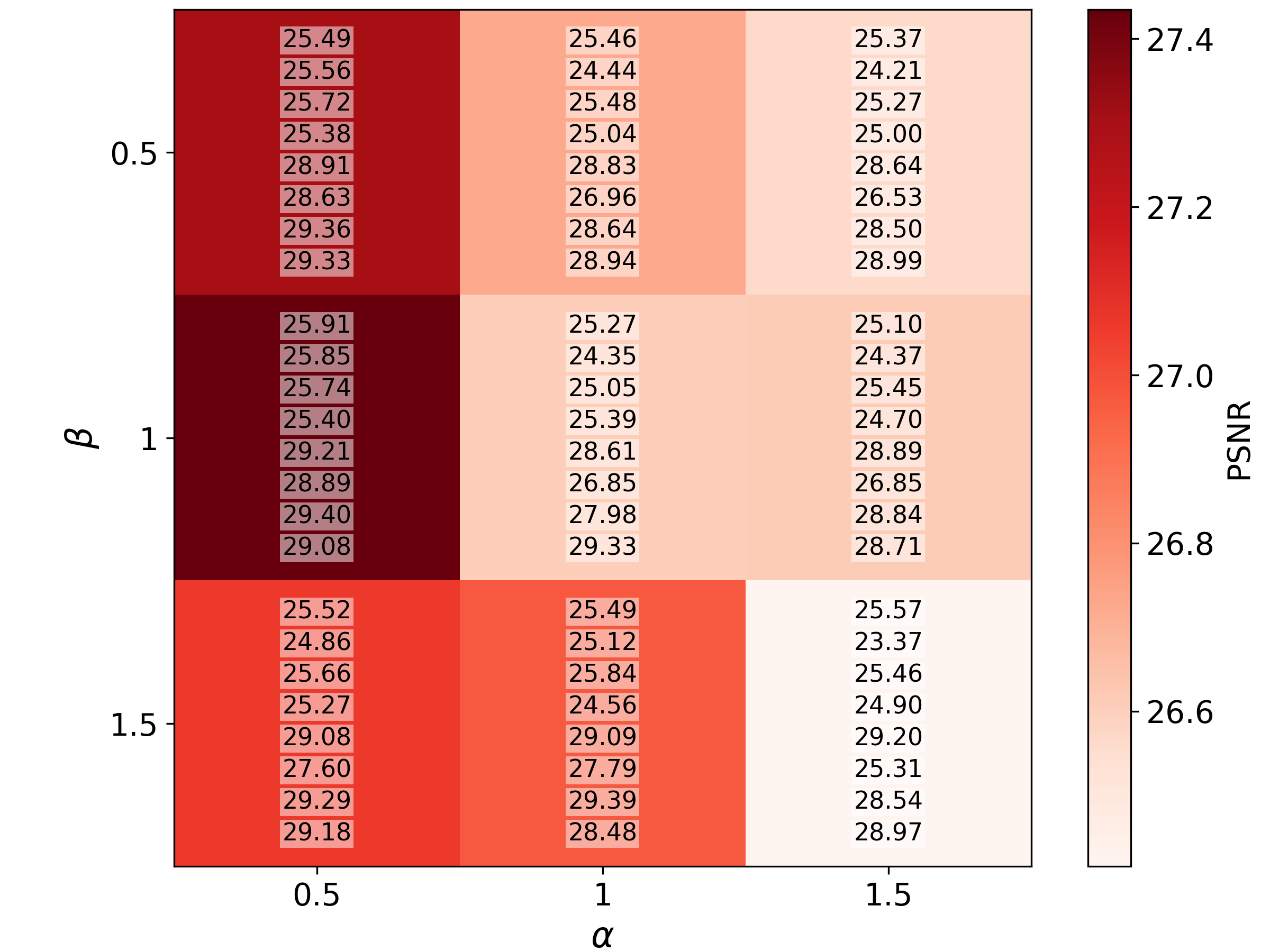}
        \caption{}
        \label{fig:FFLablation_psnr}
    \end{subfigure}
    \hfill
    \begin{subfigure}[b]{0.49\textwidth}
        \centering
        \includegraphics[width=0.9\textwidth]{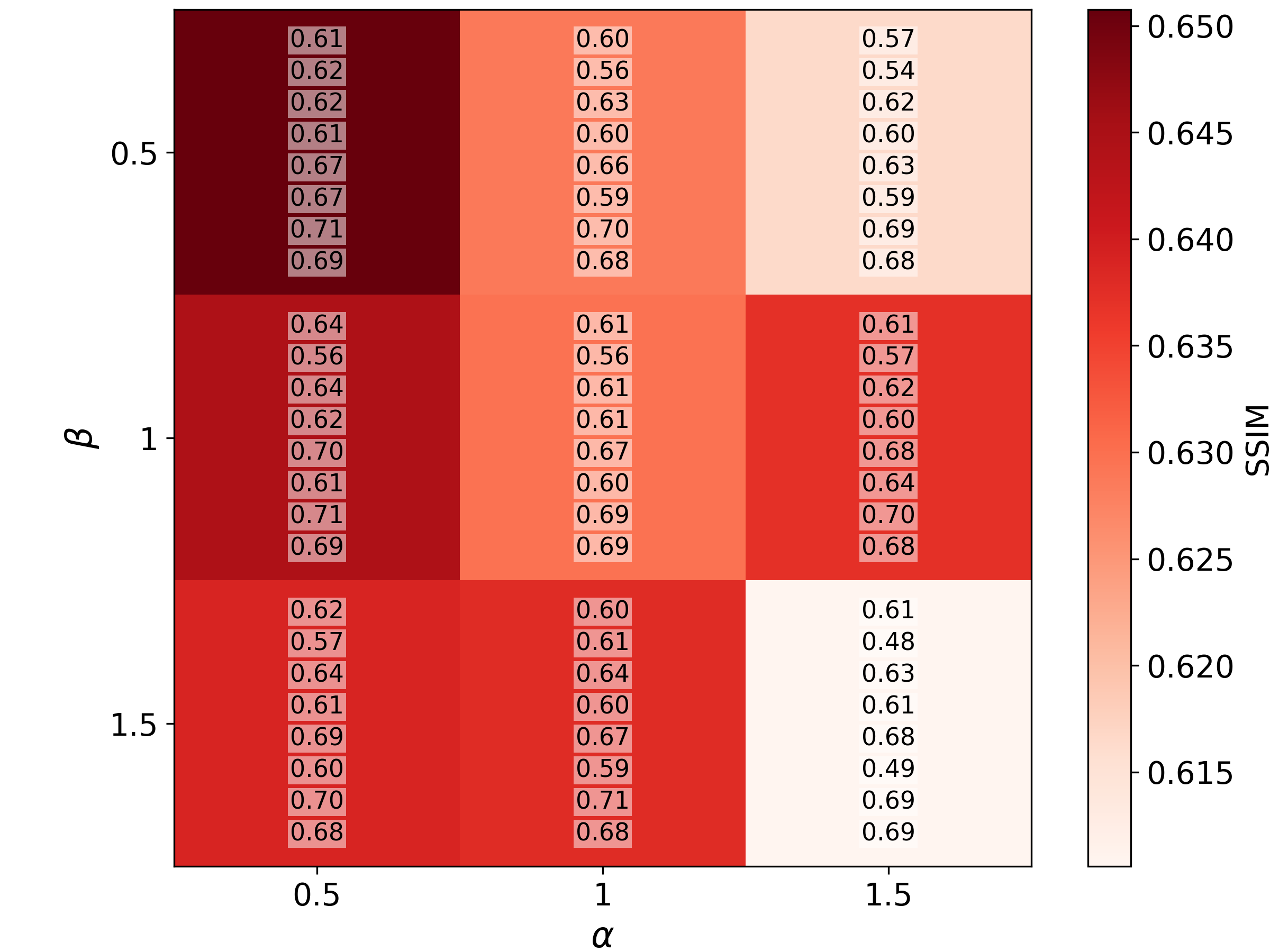}
        \caption{}
        \label{fig:FFLablation_ssim}
    
    \end{subfigure}
    
    \caption{Heatmaps with the mean values of PSNR (a) and SSIM (b) evaluated on the test dataset after training the networks using the $\mathcal{L}_{\textit{FFL}}^{\beta,\alpha}$ loss function with various combinations of the parameters $\alpha$ and $\beta$ (x and y-axis, respectively). Each cell represents the mean of $8$ values, the first $4$ corresponding to the parameter value evaluated on \(\mathcal{D}_\text{Art}^{Ts}\), and the last $4$ corresponding to the parameter value evaluated on the \(\mathcal{D}_\text{All}^{Ts}\), for each neural network in the study, MAR-DTN, pix$2$pix, custom-pix2pix, and SwinIR, respectively.}
    \label{fig:new2}
\end{figure}

\subsubsection{$\mathcal{L}_{\textit{FFL}}^{\beta,\alpha}$ Analysis}

$\beta$ and $\alpha$ were varied within the set $\{0.5,1,1.5\}$. The average value of the metrics evaluated on the \(\mathcal{D}_\text{All}^{Ts}\) set can be seen in Fig. \ref{fig:FFLablation_psnr} and Fig.  \ref{fig:FFLablation_ssim}.

As with the prior study, we observe some variability in the results with less than $2$dB in PSNR and $10\%$ in SSIM.

However, it is observed that increasing the value of alpha decreases the metrics. For the minimum $\alpha$ value, $\alpha=0.5$, the best PSNR result is for $\beta=1$, with a PSNR value of $27.81$dB. On the other hand, the mean SSIM value when $alpha=0.5$ and $\beta=1$ is $0.64$, very close to the best value, which is $0.65$. Taking this into account, we conclude that the best combination of values is $\beta=1$ and $\alpha=0.5$.


\subsubsection{Loss function comparison}
\label{sub:lossFuncResults}
Table \ref{table:comparativeResults} compares the results obtained considering different loss functions combinations and datasets. 

The $\mathcal{L}_1^{100}$ loss function achieves the best results on \(\mathcal{D}_\text{Art}\) for MAR-DTN, with a PSNR of $27.17$dB, and it is the second best result for pix$2$pix, with a PSNR of $26.31$dB. However, for both custom-pix$2$pix and SwinIR, the performance is reduced by almost $2$dB.

\begin{table*}[htb]
\centering
\caption{Comparative analysis for different networks and loss function combinations, indicated with a check mark which sum of loss functions have been used for training. For the Pix$2$Pix networks, it indicates the loss function of the generator. The dataset column indicates the dataset with which the network has been trained and evaluated; where dataset is \(\mathcal{D}_\text{All}\) then model is trained on \(\mathcal{D}_\text{All}^{Tr}\) and tested on \(\mathcal{D}_\text{All}^{Ts}\), and in case of \(\mathcal{D}_\text{Art}\) then model is trained on \(\mathcal{D}_\text{Art}^{Tr}\), and tested on \(\mathcal{D}_\text{Art}^{Ts}\). Finally, the remaining columns show the PSNR and SSIM values obtained for the test sets. Where the dataset is the \(\mathcal{D}_\text{All}\), we report both on the performance obtained on artifact slices from within the \(\mathcal{D}_\text{Art}^{Ts}\), and the mean of PSNR and SSIM on whole dataset \(\mathcal{D}_\text{All}^{Ts}\) (in parentheses). Underlined values indicate the highest performance for each network with certain loss function combinations, while highlighted values indicate the highest overall performing model across all configurations.}

\label{table:comparativeResults}
\tiny
\resizebox{\textwidth}{!}{
\begin{tblr}{
  cells = {c},
  cell{1}{1} = {c=5}{},
  cell{1}{6} = {r=2}{},
  cell{1}{7} = {c=2}{},
  cell{1}{9} = {c=2}{},
  cell{1}{11} = {c=2}{},
  cell{1}{13} = {c=2}{},
  cell{1}{15} = {c=2}{},
  hline{1,3,5,7,9,11,13,15,17,19,21} = {-}{},
  vline{6,7,9,11,13,15,17} = {-}{},
  hline{2} = {1-5,7-16}{}
}

Loss combination                                    &                                                           &                                                              &                                                          &                                                                  & \begin{sideways}Dataset\end{sideways} & MAR-DTN              &                    & pix2pix \cite{8100115}              &                    & custom-pix2pix       &                    & SwinIR \cite{liang2021swinir}               &                    & INet \cite{weng2021inet}                 &                    \\
\begin{sideways}$\mathcal{L}_1^{100}$\end{sideways} & \begin{sideways}$\mathcal{L}_\textit{SSIM}$\end{sideways} & \begin{sideways}$\mathcal{L}_\textit{MS-SSIM}$\end{sideways} & \begin{sideways}$\mathcal{L}_\textit{MSE}$\end{sideways} & \begin{sideways}$\mathcal{L}_\textit{FFL}^{1,0.5}$\end{sideways} &                                       & PSNR                 & SSIM               & PSNR                 & SSIM               & PSNR                 & SSIM               & PSNR                 & SSIM               & PSNR                 & SSIM               \\
$\checkmark$                                        &                                                           &                                                              &                                                          &                                                                  & \(\mathcal{D}_\text{Art}\)                                   & $27.17$              & $0.69$             & $26.31$              & $0.64$             & $25.24$              & $0.68$             & $25.46$              & $0.61$             & $11.61$              & $0.04$             \\
                                                    &                                                           &                                                              &                                                          &                                                                  & \(\mathcal{D}_\text{All}\)                                   & {$26.99$\\($28.97$)} & {$0.71$\\($0.73$)} & {$26.36$\\($28.7$)}  & {$0.63$\\($0.69$)} & {$26.61$\\($29.08$)} & {$0.67$\\($0.7$)}  & {$25.29$\\($28.79$)} & {$0.59$\\($0.66$)} & {$12.02$\\($12.03$)} & {$0.04$\\($0.04$)} \\
$\checkmark$                                        & $\checkmark$                                              &                                                              &                                                          &                                                                  & \(\mathcal{D}_\text{Art}\)                                   & $27.11$              & $0.69$             & $26.21$              & $0.63$             & $26.98$              & $0.71$             & $26.16$              & $0.62$             & $10.94$              & $0.08$             \\
                                                    &                                                           &                                                              &                                                          &                                                                  & \(\mathcal{D}_\text{All}\)                                   & {$\colorbox{lightgrey}{\underline{27.09}}$\\($\colorbox{lightgrey}{\underline{30.02}}$)} & {$\colorbox{lightgrey}{\underline{0.69}}$\\($\colorbox{lightgrey}{\underline{0.73}}$)} & {$26.39$\\($28.58$)} & {$0.64$\\($0.68$)} & {$\underline{27.13}$\\($\underline{29.85}$)} & {$\underline{0.68}$\\($\underline{0.70}$)} & {$24.90$\\($28.97$)} & {$0.68$\\($0.68$)} & {$12.01$\\($12.27$)} & {$0.03$\\($0.03$)} \\
$\checkmark$                                        &                                                           & $\checkmark$                                                 &                                                          &                                                                  & \(\mathcal{D}_\text{Art}\)                                   & $\colorbox{lightgrey}{27.46}$              & $\colorbox{lightgrey}{0.69}$             & $26.32$              & $0.65$             & $27.06$              & $0.67$             & $26.26$              & $0.63$             & $11.96$              & $0.01$             \\
                                                    &                                                           &                                                              &                                                          &                                                                  & \(\mathcal{D}_\text{All}\)                                   & {$27.08$\\($29.97$)} & {$0.69$\\($0.73$)} & {$26.37$\\($28.69$)} & {$0.58$\\($0.68$)} & {$27.04$\\($29.35$)} & {$0.64$\\($0.70$)} & {$\underline{26.25}$\\($\underline{29.39}$)} & {$\underline{0.63}$\\($\underline{0.67}$)} & {$12.67$\\($12.95$)} & {$0.05$\\($0.04$)} \\
$\checkmark$                                        &                                                           &                                                              & $\checkmark$                                             &                                                                  & \(\mathcal{D}_\text{Art}\)                                   & $26.94$              & $0.68$             & $26.25$              & $0.64$             & $26.55$              & $0.64$             & $26.18$              & $0.63$             & $11.70$              & $0.02$             \\
                                                    &                                                           &                                                              &                                                          &                                                                  & \(\mathcal{D}_\text{All}\)                                   & {$27.11$\\($29.89$)} & {$0.69$\\($0.72$)} & {$\underline{26.42}$\\($\underline{28.92}$)} & {$\underline{0.68}$\\($\underline{0.7}$)}  & {$26.98$\\($29.22$)} & {$0.64$\\($0.68$)} & {$25.24$\\($29.00$)} & {$0.60$\\($0.66$)} & {$11.23$\\($10.96$)} & {$0.01$\\($0.01$)} \\
                                                    &                                                           &                                                              &                                                          & $\checkmark$                                                     & \(\mathcal{D}_\text{Art}\)                                   & $26.35$              & $0.64$             & $24.03$              & $0.51$             & $26.15$              & $0.61$             & $24.58$              & $0.59$             & $9.05$               & $0.08$             \\
                                                    &                                                           &                                                              &                                                          &                                                                  & \(\mathcal{D}_\text{All}\)                                   & {$26.52$\\($29.51$)} & {$0.66$\\($0.70$)} & {$25.85$\\($28.88$)} & {$0.56$\\($0.61$)} & {$26.02$\\($29.04$)} & {$0.60$\\($0.69$)} & {$25.39$\\($29.33$)} & {$0.61$\\($0.69$)} & {$10.03$\\($10.01$)} & {$0.02$\\($0.01$)} \\
$\checkmark$                                        &                                                           &                                                              &                                                          & $\checkmark$                                                     & \(\mathcal{D}_\text{Art}\)                                   & $27.06$              & $0.69$             & $25.66$              & $0.63$             & $26.66$              & $0.60$             & $25.40$              & $0.61$             & $11.95$              & $0.03$             \\
                                                    &                                                           &                                                              &                                                          &                                                                  & \(\mathcal{D}_\text{All}\)                                   & {$26.99$\\($29.85$)} & {$0.69$\\($0.72$)} & {$25.99$\\($28.56$)} & {$0.59$\\($0.65$)} & {$26.48$\\($29.06$)} & {$0.62$\\($0.69$)} & {$25.55$\\($29.18$)} & {$0.60$\\($0.68$)} & {$11.54$\\($11.86$)} & {$0.08$\\($0.07$)} \\
$\checkmark$                                        & $\checkmark$                                              &                                                              &                                                          & $\checkmark$                                                     & \(\mathcal{D}_\text{Art}\)                                   & $27.08$              & $0.69$             & $25.66$              & $0.63$             & $26.18$              & $0.56$             & $26.42$              & $0.64$             & $9.58$               & $0.05$             \\
                                                    &                                                           &                                                              &                                                          &                                                                  & \(\mathcal{D}_\text{All}\)                                   & {$25.65$\\($28.65$)} & {$0.63$\\($0.69$)} & {$26.41$\\($28.74$)} & {$0.68$\\($0.64$)} & {$27.08$\\($28.04$)} & {$0.64$\\($0.68$)} & {$25.45$\\($28.79$)} & {$0.61$\\($0.66$)} & {$12.53$\\($12.21$)} & {$0.06$\\($0.05$)} 
\end{tblr}
}
\end{table*}

\begin{figure}[h]
\centering
\includegraphics[width=0.7\textwidth]{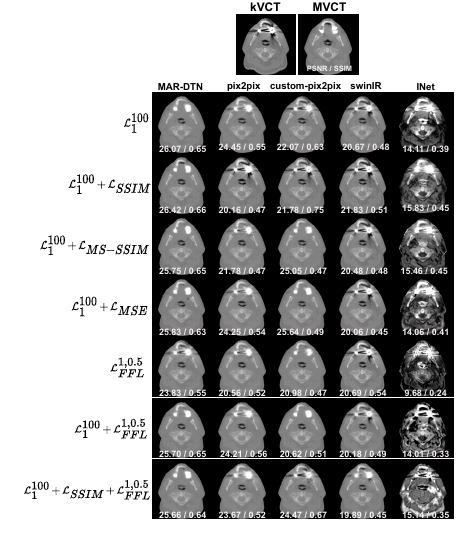}
\caption{Reconstruction of a slice with artifacts by the different models and loss functions. First row shows preprocessed kVCT and MVCT images (ground truth). First column indicates the loss function, and the following ones indicate the model used. Networks have been trained on the \(\mathcal{D}_\text{Art}\).} 
\label{fig:sliceReconstruction}
\end{figure}

Also the $\mathcal{L}_\textit{FFL}^{1,0.5}$ loss function has been tested alone. It yields less accurate results than $\mathcal{L}_1^{100}$, decreasing the PSNR value by up to $2$dB for pix$2$pix. However, custom-pix$2$pix maintains a PSNR value of $26.15$dB, competitive with the rest of the loss functions. Regarding the rest of the loss functions, the improvement in using $\mathcal{L}_{MS-SSIM}$ instead of $\mathcal{L}_{SSIM}$ stands out, especially notable in custom-pix$2$pix. The most complex loss function ($\mathcal{L}_1^{100} + \mathcal{L}_{MS-SSIM} + \mathcal{L}_\textit{FFL}^{1,0.5} $) introduces noise during training and fails to surpass the metric value achieved by simpler functions. Nonetheless, SwinIR achieves the best result, with a PSNR of $26.42$dB. An example of the reconstruction of a slice with artifacts can be seen in Fig.  \ref{fig:sliceReconstruction}.

However, the results in the \(\mathcal{D}_\text{All}\) improve when other loss functions are added to $\mathcal{L}_1^{100}$ . In the case of MAR-DTN and custom-pix$2$pix, the best combination is $\mathcal{L}_1^{100} +\mathcal{L}_{SSIM}$, reaching a PSNR value of $30.02$dB. For pix$2$pix and SwinIR, the best combination is $\mathcal{L}_1^{100} +\mathcal{L}_{MSE}$ with a PSNR of $28.92$ and $29.39$dB, respectively. 

In general, the metric values for slices with artifacts are lower when trained on \(\mathcal{D}_\text{All}\). This is consistent with having an unbalanced dataset and means that the loss functions are not entirely capable of addressing the issue of class imbalance.

On the other hand, the results obtained with INet can be found in the last column of Table \ref{table:comparativeResults}. The highest PSNR achieved is $12.67$dB for the artifact set, highlighting the architecture's inability to compete with the other architectures in the study. For the SSIM, we observe a similar behavior, with a maximum value of $8\%$. In Fig.  \ref{fig:sliceReconstruction}, the results obtained by INet can be seen. The artifacts not only have not been reduced, but they also acquire higher contrast, along with the rest of the image. However, new artifacts appear, which blur and deform other structures; for example, noticeable in the gird where streaks obtained with a combination of loss functions $\mathcal{L}_1^{100} + \mathcal{L}_{MS-SSIM} + \mathcal{L}_\textit{FFL}^{1,0.5} $.

\subsection{State-of-the-Art Comparison}

The comparison between MAR-DTN and state-of-the-art networks (pix$2$pix, custom-pix$2$pix, and SwingIR using $\mathcal{L}_1^{1}$ as loss function) shows MAR-DTN achieves the best result for the \(\mathcal{D}_\text{Art}\), with a PSNR of $26.99$dB and an SSIM of $0.69$ points. However, custom-pix$2$pix and MAR-DTN achieve the best overall result for the \(\mathcal{D}_\text{All}\), with a PSNR of $29.88$dB and an SSIM of $0.73$ points for pix$2$pix. SwinIR exhibits a decrease of up to $0.76$dB for PSNR and $0.07$ points for SSIM across the \(\mathcal{D}_\text{All}\), with a larger decrease observed within the \(\mathcal{D}_\text{Art}\), reaching a difference of $1.7$dB compared to MAR-DTN. Table \ref{table:sotaComputationalResources} presents a comparison between the performance time and complexity of the networks.

\begin{table}[h]
\centering
\caption{Comparison of trainable parameters, number of multiplications and additions (MACs), training time computed for the \(\mathcal{D}_\text{All}\) in $1$ epoch and patient reconstruction time (in this case $170$ slices) for state-of-the-art methods under study.}
\label{table:sotaComputationalResources}
\begin{tblr}{
  cells = {c},
  hline{1,7} = {-}{0.12em},
  hline{2} = {-}{},
}
Network        & {Parameters \\(M)} & {MACs \\(G)} & {Training time\\(s)} & {Patient reconstruction\\time (s)} \\
MAR-DTN         & $1.882$            & $116.686$     & $65.32$              & $3.56$                               \\
pix2pix        & $54.413$           & $77.99$      & $80.02$              & $3.75$                               \\
custom-pix2pix & $4.646$            & $123.277$     & $67.42$                 & $4.25$                               \\
SwinIR         & $1.614$            & $425.034$     & $2,774.76$           & $47.27$   \\
INet         & $2.96$            & $896.31$     & $807.38$           & $5.31$  
\end{tblr}
\end{table}


\subsection{Clinical evaluation}
Initial feedback from clinicians indicates that the quality of the MVCT images generated through our proposed method is highly regarded. Clinicians have noted that synthetic MVCT images exhibit excellent contrast for both soft tissues and bones, which is essential for accurate diagnosis and treatment planning in clinical practice. These qualitative observations suggest promising outcomes in terms of image quality and clinical utility, laying a strong foundation for further quantitative evaluation and validation studies in the future.

\section{Discussion and Conclusion}
In this study, we compared our proposed domain transformation methodology with some state-of-the-art methods, where kVCT images serve as input and MVCT images as output. Our results demonstrate that a lightweight model like MAR-DTN can effectively reduce artifacts with the appropriate combination of loss functions, even with a reasonable dataset size. The performance of the models is evaluated on two datasets: $\mathcal{D}_\text{Art}$, which contains only images with artifacts, and $\mathcal{D}_\text{All}$ includes both artifact-affected and non-affected images. 

Numerous combinations of loss functions were tested, though only a select few are presented in Table \ref{table:comparativeResults} due to space constraints. Consequently, a deliberate choice was made to include those combinations yielding more promising results within the allocated space.

As we compare the performance of models trained on $\mathcal{D}_\text{All}$, MAR-DTN shows the best performance in several cases, especially with the combination of $\mathcal{L}_1^{100} + \mathcal{L}_{SSIM}$ when tested on \(\mathcal{D}_\text{All}^{Ts}\), achieving the highest PSNR of 30.02 dB and a high SSIM of 0.73 on over all patient volume, in addition to that when tested on \(\mathcal{D}_\text{Art}^{Ts}\) still achieves competitive results. Model pix2pix and custom-pix2pix show similar performance, with custom-pix2pix slightly outperforming pix2pix in most cases. custom-pix2pix performs best on $\mathcal{D}_\text{All}$ with the loss combination of $\mathcal{L}_1^{100} + \mathcal{L}_{SSIM}$ and pix2pix show fair performance using $\mathcal{L}_1^{100} +\mathcal{L}_{MSE}$. SwinIR exhibits decent performance but is generally outperformed by MAR-DTN, particularly in terms of PSNR. However, it shows competitive SSIM values. Model INet performs the worst among all models, with significantly lower PSNR (max 12.67 dB) and SSIM (max 0.08) values, highlighting its inability to effectively reduce artifacts or maintain structural similarity.

Furthermore, as we compare the performance of models trained on $\mathcal{D}_\text{Art}$, MAR-DTN achieves the highest performance on this dataset with the combination of $\mathcal{L}_1^{100} + \mathcal{L}_{MS-SSIM}$, achieving a PSNR of 27.46 dB and an SSIM of 0.69 when tested on \(\mathcal{D}_\text{Art}^{Ts}\). Overall, MAR-DTN performs better than all other models across various loss combinations, particularly on the $\mathcal{D}_\text{All}$ dataset. Model pix2pix and custom-pix2pix show similar PSNR and SSIM values, typically around 26-27 dB for PSNR and 0.64-0.68 for SSIM, depending on the loss function combination used. Model custom-pix2pix slightly outperforms pix2pix in most of the combinations. Model SwinIR performs reasonably well, achieving PSNR values around 25-26 dB and SSIM values around 0.64-0.67, depending on the loss function combination but it is outperformed by MAR-DTN in most combinations. INet shows the poorest performance on $\mathcal{D}_\text{Art}$. 
We can conclude that it is not capable of eliminating artifacts using the loss functions in this study, falling far behind its competitors. What is achieved, however, is an increase in contrast between different bone and muscle structures. Nevertheless, it also introduces new artifacts, which hinder the correct evaluation of the images. It is important to note that INet's initial goal is image segmentation, not image generation. Additionally, INet performs better with low-resolution images, making it less appropriate for our dataset.

Despite achieving satisfactory results, it is worth noting that our study's considered dataset is relatively small. Nevertheless, our approach demonstrates significant potential, as evidenced by MAR-DTN's robust performance metrics achieved across various network architectures and loss functions. To further enhance the impact of our findings, we plan to incorporate systematic qualitative evaluations by clinical staff. Additionally, we aim to expand our dataset to include more types of artifacts across different body regions. These steps will provide deeper insights and potentially lead to even more improved results, reinforcing the efficacy and applicability of our methodology in broader contexts. Moreover, our future work aims to develop a generalized model for the entire body. This extension will significantly broaden the applicability and robustness of our approach, paving the way for more comprehensive and versatile artifact management in medical imaging.

\subsubsection{Acknowledgements} This work is supported by the Italian Ministry of Health (Ricerca Corrente). APM and BSA acknowledge financial support by the Spanish Ministerio de Economía y Competitividad and European Regional Development Fund, MCIN/AEI/ 10.13039/501100011033 and by “ERDF A way of making Europe”. Xunta de Galicia funded research under Research Grant No. 2021-PG036 and the Spanish Ministerio de Ciencia e Innovación MCIN/AEI/10.13039/501100011033 through the Industrial Doctorates Grant.



%
%

\bibliographystyle{splncs04}
\bibliography{bibliography}

\end{document}